# New Constraints on the Major Neutron Source in Low-mass AGB Stars

Nan Liu[1,2], Roberto Gallino[3], Sergio Cristallo[4,5], Sara Bisterzo[6]

Andrew M. Davis[7,8,9], Reto Trappitsch[10], Larry R. Nittler[11]

[1]Laboratory for Space Sciences and Physics Department, Washington University in St. Louis, St. Louis, MO 63130, USA; nliu@physics.wustl.edu

[2]McDonnell Center for the Space Sciences, Washington University in St. Louis, St. Louis, MO 63130, USA

[3]Dipartimento di Fisica, Università di Torino, Torino 10125, Italy

[4]INAF-Osservatorio Astronomico d'Abruzzo, Teramo 64100, Italy

[5]INFN-Sezione di Perugia, Perugia 06123, Italy

[6]INAF-Osservatorio Astrofisico di Torino, Pino Torinese 10025, Italy

[7]Department of the Geophysical Sciences, The University of Chicago, Chicago, IL 60637, USA

[8]Chicago Center for Cosmochemistry, Chicago, IL 60637, USA

[9]The Enrico Fermi Institute, The University of Chicago, Chicago, IL 60637, USA

[10]Nuclear and Chemical Sciences Division, Lawrence Livermore National Laboratory, Livermore, CA 94550, USA

[11]Department of Terrestrial Magnetism, Carnegie Institution for Science, Washington, DC 20015, USA

## ABSTRACT

We compare updated Torino postprocessing asymptotic giant branch (AGB) nucleosynthesis model calculations with isotopic compositions of mainstream SiC dust grains from low-mass AGB stars. Based on the data-model comparison, we provide new constraints on the major neutron source, $^{13}C(\alpha,n)^{16}O$ in the He-intershell, for the s-process. We show that the literature Ni, Sr, and Ba grain data can only be consistently explained by the Torino model calculations that adopt the recently proposed magnetic-buoyancy-induced $^{13}C$-pocket. This observation provides strong support to the suggestion of deep mixing of H into the He-intershell at low $^{13}C$ concentrations as a result of efficient transport of H through magnetic tubes.

## 1. INTRODUCTION

The theory for the formation of heavy nuclides (nuclides beyond the iron peak elements) was laid out by Burbidge et al. (1957) and Cameron (1957) more than a half-century ago. Based on a systematic analysis of the solar system abundance distribution, they proposed three distinct processes for synthesizing the heavy nuclides, including slow (s) neutron-capture process, rapid (r) neutron-capture process, and proton (p) capture process, with the former two being responsible for making most of the heavy nuclides. These terms are still being used today, although additional intermediate neutron-capture processes such as the i-process (Cowan & Rose 1977) have been proposed to also occur in stars (Herwig et al. 2011).

The main s-process[1] occurs in thermally pulsing asymptotic giant branch (AGB) stars, the final evolutionary stage of low-to-intermediate mass stars. Figure 1 illustrates the evolution of a 3 $M_\odot$, 1.5 $Z_\odot$[2] AGB star predicted by FRUITY model calculations (Cristallo et al. 2011). The star can be divided into three main regions: an innermost zone corresponding to the partially degenerate CO core, inside the magenta line; an external zone corresponding to the H-rich convective envelope, outside the black line; and a thin region ($10^{-2}-10^{-3}$ $M_\odot$) between the magenta and black lines, called the He-intershell. The red and blue lines show where maximum energy release from H-burning and within the He-intershell takes place, respectively. The blue line shows a plateau during each interpulse period, where the main energy release in the He-intershell is from the $^{13}C(\alpha,n)^{16}O$ reaction. The base of the H-rich envelope is radiative between the black and red lines. Hydrogen burning is activated for most of the time, on the top of the growing inert He-intershell. The He-intershell is heated and compressed until the temperature and density at its bottom are high enough to trigger thermonuclear runaways by He-burning (Thermal Pulse, TP, Iben & Renzini 1983). Because of the surplus of energy locally released, the whole He-intershell becomes unstable and convects. As a consequence of the energy release and expansion of the He-shell, the overlying layers, including the H-shell, are pushed to larger radii (e.g., numbered TPs in Fig. 1). At the quenching of a TP, the He-shell returns to radiative conditions; conversion of a significant fraction of the $^{4}$He in the He-shell into $^{12}$C is responsible for the jumps in size of the CO core (magenta line) after each TP. After a few hundred years, if the H-shell has been sufficiently lifted by the TP, the convective envelope penetrates the underlying region and brings newly synthesized materials to the surface. This process is called third dredge-up (TDU). Consequently, the convective envelope becomes more and more C-rich. As soon as the temperature becomes sufficiently high, H burning is reactivated in the H-shell. The time elapsed from the maximum penetration of the convective envelope and the hydrogen reignition is a few thousand years. An AGB star suffers mass loss from the surface in the form of stellar winds, losing 50% to 85% of its mass during the AGB phase (depending on its initial mass). For a 3 $M_\odot$ star, its TP-AGB phase lasts for about 2.5 Ma (Fig. 1). When the mass of the H-rich envelope becomes very small, the star leaves the AGB phase toward the White Dwarf cooling sequence.

Technetium is an element that does not have any stable isotopes. One of its radioactive isotopes, $^{99}$Tc ($t_{1/2}$ =0.2 Ma), sits along the s-process path while being shielded from the paths of p- and r-process. The detection of Tc in AGB stars, therefore, clearly points to in situ production of $^{99}$Tc by the s-process during the previous evolution of the stars, thus confirming AGB stars as the stellar site for the s-process nucleosynthesis (Merrill 1952). Guided by observed heavy-element abundances on the surface of low-mass AGB stars, subsequent numerical modeling of AGB stellar nucleosynthesis found that $^{13}C(\alpha,n)^{16}O$ acts as the major neutron source and $^{22}Ne(\alpha,n)^{25}Mg$ as the minor neutron source for the s-process in low-mass AGB stars (Gallino et al. 1990). The former occurs radiatively at neutron energies of about 8 keV between TPs, providing neutrons with a neutron density of ~$10^{7}-10^{8}$ cm$^{-3}$ on the timescale of 5−20 ka. In contrast, the latter occurs convectively at neutron energies of about 23 keV during TPs, providing a short neutron burst with a neutron density up to $10^{9}-10^{10}$ cm$^{-3}$ on the timescale of a few years (Gallino et al. 1998). The short neutron burst released by the minor neutron source $^{22}$Ne mainly controls production of nuclides affected by branch points along the s-process path (e.g., Käppeler et al. 2011).

---

[1] Unless noted otherwise, the s-process hereafter refers to the main s-process nucleosynthesis.

[2] The solar metallicity, $Z_\odot$, is defined as the protosolar abundance by mass of elements heavier than He, corresponding to a value of 0.014 (Lodders 2003) in FRUITY stellar model calculations.

s-Process nucleosynthesis in AGB stars plays a fundamental role in our understanding of the variety of nucleosynthetic processes responsible for producing the heavy nuclides, especially in the case of the r-process (e.g., Cowan et al. 1991; Goriely & Arnould 2001; Pearson et al. 2005) and the weak s-process (Pignatari et al. 2010), as contributions to solar system abundances from processes other than the s-process are calculated by subtracting modeled s-process abundances from solar system abundances (e.g., Arlandini et al. 1999; Bisterzo et al. 2014). Parameters in AGB stellar models as well as nuclear uncertainties, however, both contribute to uncertainties in model predictions for the two s-process neutron sources and thus for the s-process (Käppeler et al. 2011). While uncertainty in the minor $^{22}$Ne neutron source arises mainly from experimental uncertainty in measuring the $^{22}$Ne$(\alpha,n)^{25}$Mg reaction rate (Wiescher et al. 2012; Bisterzo et al. 2014; Massimi et al. 2017), uncertainty in the major $^{13}$C neutron source results from a more fundamental problem in stellar models, namely the source of $^{13}$C for the reaction. The formation of $^{13}$C in the He-intershell requires partial mixing of H from the formal convective envelope border into the He-intershell, which contains copious $^{12}$C produced by the $3\alpha$ reaction, in order for $^{12}$C$(p,\gamma)^{13}$N$(\beta^+\nu)^{13}$C to take place in the initially H-free He-intershell. The region of the He-intershell that contains $^{13}$C between TPs is generally referred to as the "$^{13}$C-pocket." Partial mixing of H from the envelope border to the He-intershell can be induced by overshoot (Freytag et al. 1996; Herwig et al. 1997; Straniero et al. 2006; Cristallo et al. 2009; Battino et al. 2016) and gravity waves (Denissenkov & Tout 2003) as a result of the turbulent motion along this boundary. In AGB stars, additional mixing induced by rotation also needs to be considered (Herwig et al. 2003; Piersanti et al. 2013), as well as mixing related to the presence of magnetic fields (Nucci & Busso 2014; Trippella et al. 2016). Given that the way in which each of these physical instabilities should be treated is already a highly debated matter among stellar modelers; the fact that, in reality, these mixing mechanisms should operate simultaneously and likely interact with each other exacerbates the problem here. The formation of $^{13}$C depends essentially on the H mixed into the He-intershell. Since the mixing depth, density, and distribution of H within the $^{13}$C-pocket are all uncertain (Liu et al. 2014a, 2014b, 2015), the formation of the major $^{13}$C neutron source therefore remains one of the major uncertainties in modeling s-process nucleosynthesis in AGB stars (Lugaro et al. 2003; Liu et al. 2014a; Buntain et al. 2017).

Astronomical observations of the abundance distribution patterns of s-process elements for AGB stars played a fundamental role in identifying the two s-process neutron sources (Busso et al. 2001). Such observational data for s-process elemental abundances, however, are not sufficiently sensitive to the different parameters for the $^{13}$C-pocket adopted in AGB stellar models (Buntain et al. 2017), because the s-process essentially modifies the abundances of individual isotopes rather than individual elements. The abundances of individual nuclides, although unavailable from astronomical observations with the required precision, can in fact be obtained by analyzing isotopic compositions of ancient AGB dust grains extracted from primitive meteorites in laboratories on Earth. Dust grains condensed around dying stars are ejected into the interstellar medium by stellar winds and explosions. A fraction of these stardust grains survived destruction processes in the interstellar medium and in the early solar protoplanetary disk, and later got incorporated into asteroidal bodies and delivered to Earth in meteorites. Since these stardust grains are essentially stellar materials whose isotopic signatures are affected by both Galactic Chemical Evolution (GCE) and stellar nucleosynthesis, some, if not all, of them can be identified via their exotic isotopic signatures with respect to the range of isotopic compositions of solar system objects (Zinner et al. 1987; Lewis et al. 1990). These anomalous grains identified in primitive meteorites are called presolar grains.

Silicon carbide (SiC) is a refractory mineral phase and condenses in C-rich (C/O>1) stellar environments (Lodders & Fegley 1995). Among the variety of presolar minerals identified so far, presolar SiC grains have been most extensively studied for their isotopic and structural compositions (see recent reviews by Zinner 2014 and Nittler & Ciesla 2016). Such detailed analyses have demonstrated that the majority of presolar SiC grains (>90%), known as mainstream (MS) SiC grains, came from low-mass AGB stars and contain heavy trace elements with $s$-process isotopic signatures (Nicolussi et al. 1997, 1998; Lugaro et al. 2003). Analysis of heavy element isotopic compositions of MS SiC grains with precision on the order of 1−10% (e.g., Savina et al. 2003; Stephan et al. 2016), therefore, provides a unique opportunity to investigate the $s$-process nucleosynthesis in AGB stars at a level of detail that cannot be achieved by current astronomical observations.

Given that the $s$-process occurs in a relatively low neutron-density environment in AGB stars, the $s$-process neutron flow reaches steady state between magic neutron numbers, where the product of the Maxwellian averaged stellar ($n,\gamma$) cross section of a nuclide, $<\sigma>$, and its corresponding abundance, $N_s$, remains almost constant (Fig. 2 of Käppeler et al. 2011). With the exception of those affected by branch points, the abundances of nuclides sitting between magic neutron numbers are, therefore, inversely proportional to their stellar neutron-capture cross sections, and consequently are almost insensitive to the parameters for the $^{13}$C-pocket adopted in AGB stellar models. On the other hand, the neutron-magic nuclides around masses 88, 138, and 208 (corresponding to magic neutron numbers 50, 82, and 126, respectively) act as bottlenecks for the neutron-capture flow along the $s$-process path due to their extremely small neutron-capture cross sections, resulting in three distinct steps in the $<\sigma>N_s$ curve (Fig. 2 of Käppeler et al. 2011). Thus, the abundances of nuclides with magic numbers of neutrons (with respect to the abundances of adjacent $s$-process nuclides) are extremely sensitive to the $^{13}$C-pocket adopted in AGB models. Astronomical observations for $s$-process elemental abundances, therefore, adopt the indices of $ls$ (the total abundance of light $s$-process elements Kr, Sr, Y, Zr at the first $s$-process peak) and $hs$ (the total abundance of heavy $s$-process elements Ba, La, Ce, Nd, Sm at the second $s$-process peak) to infer the total amount of $^{13}$C produced within the $^{13}$C-pocket for an AGB star (e.g., Busso et al. 1999). However, the bottlenecks sitting at the first and second $s$-process peaks actually correspond specifically to the individual nuclides $^{86}$Kr, $^{88}$Sr, $^{89}$Y, $^{90}$Zr, and $^{138}$Ba, $^{139}$La, $^{140}$Ce, $^{141}$Pr, $^{142}$Nd, $^{144}$Sm, respectively. As a result, the specific productions of $^{88}$Sr and $^{138}$Ba relative to those of the adjacent pure $s$-process nuclides depend strongly on the $^{13}$C mass extension, $^{13}$C mass fraction, and the $^{13}$C distribution profile adopted in AGB models (Liu et al. 2014a, 2015). Although astronomical observations currently cannot achieve such detailed isotope measurements, correlated $^{88}$Sr/$^{86}$Sr and $^{138}$Ba/$^{136}$Ba ratios have been obtained in ~50 MS SiC grains with uncertainties on the order of 10% (Liu et al. 2015; Stephan et al. 2018), and can be used to investigate the fine structure of the $^{13}$C-pocket in great detail.

In this study, we will derive new constraints on the $^{13}$C-pockets in AGB stars by comparing presolar MS grain data to Torino AGB nucleosynthesis model predictions. Compared to our previous studies (Liu et al. 2014a, 2014b, 2015), three major updates are included here. (1) In addition to the correlated $^{88}$Sr/$^{86}$Sr and $^{138}$Ba/$^{136}$Ba ratios (Liu et al. 2015), we also consider recently published high-precision Ni isotope ratios of MS SiC grains (Trappitsch et al. 2018). (2) With respect to the older FRANEC code (Straniero et al. 2003) adopted in our previous studies, Torino nucleosynthesis calculations presented here were conducted based on physical quantities extracted from new FRUITY stellar models (Cristallo et al. 2009, 2011). (3) In addition to the Exponential and Flat $^{13}$C-pocket profiles investigated in our previous studies, we also tested the recently proposed $^{13}$C-pocket profile based on the consideration of magnetic buoyancy (Trippella et al. 2016; Palmerini et al. 2018).

## 2. DATA and MODELS

### 2.1. Grain Data

The isotopic data used in this study include Ti (Gyngard et al. 2018), Ni (Trappitsch et al. 2018), Sr (Liu et al. 2015; Stephan et al. 2018), Zr (Nicolussi et al. 1997; Lugaro et al. 2003; Barzyk et al. 2007), and Ba (Liu et al. 2014a, 2015; Stephan et al. 2018) isotope ratios in presolar SiC grains. All Sr, Zr, and Ba isotope data except for those from Stephan et al. (2018) were obtained with the CHARISMA instrument using Resonance Ionization Mass Spectrometry (RIMS) at Argonne National Laboratory (Savina et al. 2003). All Ni isotope data and the Sr and Ba data from Stephan et al. (2018) were obtained with a new generation of RIMS instrument, CHILI, recently built at the University of Chicago (Stephan et al. 2016). All Ti isotope data from Gyngard et al. (2018) were obtained with a NanoSIMS (Nanometer-scaled Secondary Ion Mass Spectrometry) instrument (e.g., Hoppe 2016) at Washington University, St. Louis. Except for all grains in the Nicolussi et al. study and a few grains in the Liu et al. studies, the other presolar grains were all classified as MS SiC grains based on their C and Si isotope ratios obtained by NanoSIMS. Unclassified grains are assumed to be MS grains given their dominant population (>90%) among presolar SiC grains (e.g., Nittler & Ciesla 2016). All the isotope ratios are expressed as $\delta$ values[3] with $2\sigma$ uncertainties, and the reference isotopes are $^{48}$Ti, $^{58}$Ni, $^{86}$Sr, $^{94}$Zr, and $^{136}$Ba for Ti, Ni, Sr, Zr, and Ba, respectively. Note that the $\delta^{64}$Ni values reported in Trappitsch et al. (2018) were incorrectly normalized to the measured $^{64}$Ni/$^{58}$Ni ratio in stainless steel instead of the terrestrial SiC standard (the rest of the Ni isotope ratios were correctly normalized). The corrected $\delta^{64}$Ni values using the measured $^{64}$Ni/$^{58}$Ni ratio on the SiC standard are used here and a corrigendum to Trappitsch et al. (2018) is in preparation. Also, three of the eight MS grains measured by Stephan et al. (2018) had $\delta^{135}$Ba > −400‰, likely caused by solar system Ba contamination (Liu et al. 2014a), and are therefore not included for data-model comparison here. We did not consider the Fe isotopic compositions of MS SiC grains reported by Trappitsch et al. (2018) for constraining the $^{13}$C-pocket parameters, because many grains had close-to-solar Fe isotopic compositions, indicating that severe Fe contamination occurred on the Murchison meteorite parent body and/or in the laboratory.

### 2.2 Torino AGB Models

Torino postprocessing AGB models were described by Gallino et al. (1998) in detail. In addition to recent updates reported by Bisterzo et al. (2014), the Torino postprocessing nucleosynthesis results presented in this study were calculated based on physical quantities extracted from new full evolutionary FRUITY stellar models for C-rich AGB stars of 1.5−3 $M_{\odot}$ and 0.5−1.5 $Z_{\odot}$ (Cristallo et al. 2009). Compared to the FRANEC stellar models (Straniero et al. 2003) used in previous Torino nucleosynthesis calculations, the FRUITY stellar models predict higher third dredge-up (TDU) efficiencies (Straniero et al. 2006; Cristallo et al. 2009), lower surface temperatures, and enhanced mass loss rates (Cristallo et al. 2007), which have been included in the updated Torino postprocessing code and were recently reported in the study of Trappitsch et al. (2018). Torino models adopted the solar abundances from Lodders et al. (2009) with an exception for the C isotopes ($^{12}$C/$^{13}$C=25), which are modified by deep mixing during the Red Giant phase. Also, for heavy elements the Torino code used meteoritic abundances rather than an average of meteoritic and photospheric abundances, which Lodders et al. (2009) used for some elements in their solar system abundance table. As a result, the solar metallicity adopted in Torino models corresponds to 0.0139, and is therefore almost identical to the value (0.014 from Lodders 2003) adopted in FRUITY stellar models. The reaction rates are mainly based on KADoNiS (Karlsruhe Astrophysical Database of Nucleosynthesis in Stars version v0.3, Dillmann et al. 2009)[4], but different MACS values are adopted for a number of nuclides, which are described in detail by Bisterzo et al. (2015). Note that the reaction rates related to the productions of Ni, Sr, Zr, and Ba isotopes are the same as those used in our previous studies (Liu et al. 2014a, 2014b, 2015; Trappitsch et al. 2018).

---

[3] The δ-notation is defined as the deviation in parts per thousand of the isotope ratio measured in a grain relative to that in a terrestrial standard.

As explained in Liu et al. (2014a, 2014b, 2015), the $^{13}$C mass extension (i.e., pocket mass, corresponding to the range of the x-axis in Fig. 2), $^{13}$C mass fraction (corresponding to the y-axis in Fig. 2), and $^{13}$C distribution profile can all be treated as free parameters due to uncertainties in modeling the formation of the $^{13}$C-pocket. In this study, we adopt three types of $^{13}$C profiles for use in the updated Torino model calculations, labeled as Exponential, Flat, and Trippella $^{13}$C profiles in Fig. 2. As shown by Gallino et al. (1998) and Trippella et al. (2016), the H profiles are well correlated with the corresponding $^{13}$C profiles in these three types of $^{13}$C-pockets because of their low $^{13}$C concentrations and the resultant low $^{14}$N productions. The Exponential $^{13}$C profile was implemented in the Torino models using a three-zone scheme (Zone-I, Zone-II, and Zone-III) with a total pocket mass of $1 \times 10^{-3}$ $M_\odot$ (Gallino et al. 1998). The $^{13}$C mass fraction of the Flat $^{13}$C profile corresponds to that of Zone-II in the Exponential profile. The parameters for these two $^{13}$C profiles are given in Table 2 of Liu et al. (2015). On the other hand, a new $^{13}$C profile (solid maroon line in Fig. 2) was recently derived by considering magnetic buoyancy as the mechanism responsible for transporting protons into the He-intershell (Trippella et al. 2016). This $^{13}$C profile was implemented here by approximating it using a 12-zone scheme (dashed maroon line in Fig. 2a) by averaging the amount of $^{13}$C contained within each zone. Moreover, $^{14}$N acts as a neutron poison in the $^{13}$C-pocket, which consumes neutrons via $^{14}$N$(n,p)^{14}$C. Note that the $^{14}$N/$^{13}$C ratio is fixed at ~1/30 for the Exponential and Flat $^{13}$C-pockets, because it was considered that a small amount of protons are injected into the top layers of the $^{12}$C-rich He-intershell so that only a negligible amount of $^{14}$N can form via $^{13}$C$(p,\gamma)^{14}$N (Gallino et al. 1998 and references therein). For the Trippella pocket, transport of protons into the He-intershell by magnetic flux tubes results in a deep penetration of protons at low $^{13}$C mass fractions, and, consequently, the protons are exclusively captured by $^{12}$C to produce $^{13}$C with little $^{14}$N (Trippella et al. 2016). Note that for all the three types of $^{13}$C-pockets, the pocket mass does not vary during the evolution of an AGB star in updated Torino postprocessing models, although FRUITY stellar models predict that the $^{13}$C-pocket shrinks with increasing TPs (Cristallo et al. 2009).

Historically, Torino model predictions in a specific *case* correspond to model calculations by adopting the Exponential $^{13}$C-pocket with a certain amount of $^{13}$C, which can be varied by scaling the $^{13}$C mass fractions simultaneously in all the three zones down (D) or up (U) by a multiplicative factor. Here, we follow the original definition of a *case* and extend its usage also for the Flat and Trippella pockets. Figure 2a shows the Exponential and Flat pockets in *ST case*, which is so-named for historical reasons (Arlandini et al. 1999), and the Trippella $^{13}$C-pocket in *Original case*. We further scaled the $^{13}$C mass fraction[5] by different multiplicative factors to investigate its effects on model predictions for isotope ratios of the

---

elements of interest in this study. In addition, their pocket masses were also decreased or increased by different factors (Fig. 2b) to investigate the corresponding effects.

## 3. RESULTS

According to Liu et al. (2015), the $\delta^{88}$Sr and $\delta^{138}$Ba values (hereafter Sr-Ba data) that were simultaneously obtained in MS SiC grains can be explained by the old Torino nucleosynthesis calculations by adopting relatively large $^{13}$C-pockets ($\geq 1 \times 10^{-3}$ $M_{\odot}$) and low $^{13}$C mass fractions. A subsequent study of Ni isotopic compositions of MS SiC grains (Trappitsch et al. 2018), however, found that these $^{13}$C-pocket parameters cannot successfully reproduce the slopes in the Ni three-isotope plots based on comparison with updated Torino nucleosynthesis calculations. Given differences between the FRANEC and FRUITY stellar models, it thus remains unclear whether the mismatch for the Ni isotope ratios observed by Trappitsch et al. (2018) arises from the updated stellar models.

To investigate this problem, we will compare Sr-Ba and Ni isotope data with updated Torino model calculations by adopting varying $^{13}$C-pockets. Since the Sr-Ba isotope data have large uncertainties due to their relatively low abundances in MS grains, we calculated the 2D joint histogram (10×10 pixels in x- and y-axis) for the Sr-Ba data to better illustrate the distribution of the grains in Fig. 3. The pixels in greenish blue highlight the grain-concentrated region, and are outlined by a purple box, which overlaps with all but four grains within the 2σ uncertainties. For the data-model comparison shown below, the MS grain data will be compared to AGB model predictions during the C-rich phase (e.g., symbols in Fig. 4), because equilibrium condensation calculations predict that SiC grains can only condense when the C/O ratio exceeds unity in AGB stellar winds (Lodders & Fegley 1995).

We first compare the Sr-Ba and Ni isotope data with the updated Torino AGB models for a 3 $M_{\odot}$, 1.5 $Z_{\odot}$ AGB star by adopting Flat $^{13}$C-pockets (Fig. 4). The K94 rate in Fig. 4 refers to the lower limit of the recommended value for the $^{22}$Ne($\alpha$,n)$^{25}$Mg reaction rate from Käppeler et al. (1994). According to a recent reevaluation (Longland et al. 2012) of the rate measurements by Jaeger et al. (2001), we reduced the K94 rate by a factor of two for the $^{22}$Ne($\alpha$,n)$^{25}$Mg reaction in the models presented in Fig. 4. In general, the updated Torino models that are constrained by the Sr-Ba grain data, i.e., that fall within the purple box in Fig. 4a (corresponding to the most probable grain compositions), are in good agreement with the constrained old Torino models reported by Liu et al. (2015). The best match with the grain data in Fig. 4a corresponds to 3 $M_{\odot}$, 1.5 $Z_{\odot}$ AGB model predictions that adopt a Flat $^{13}$C-pocket of 2×10$^{-3}$ $M_{\odot}$ in the ST case. In comparison, Liu et al. (2015) showed that with a Flat $^{13}$C-pocket of the same pocket mass, the best match with the grain data was given by 2 $M_{\odot}$, 0.5 $Z_{\odot}$ AGB model predictions in the D3 case[6]. Although the constrained amount of $^{13}$C in the latter is a factor of three lower than that in the former, the corresponding stellar metallicity is also a factor of three lower, thus meaning the same neutron-to-seed ratio, i.e., the same ratio of the amount of $^{13}$C to the stellar metallicity. The difference between the constrained amounts of $^{13}$C can, therefore, be well explained by the fact that the s-process efficiency is a linear function of the neutron-to-seed ratio.

Figure 4 clearly points out a dilemma faced by the model predictions for the Sr-Ba and Ni isotope ratios when adopting the Flat $^{13}$C profile to explain the grain data. Although the grain-concentrated region, i.e., the purple box, in Fig. 4a can be fairly well matched by choosing a $^{13}$C-pocket mass of 2×10$^{-3}$ $M_{\odot}$ in the ST case, this case yields much shallower slopes with respect to the Ni isotope data in Figs. 4b and 4c. On the other hand, the D2 case with a pocket mass of $1 \times 10^{-3}$ $M_\odot$ lies closest to the grain data in Figs. 4b and 4c but barely overlaps with any of the Sr-Ba grain data in Fig. 4a. Although a better match to the Sr-Ba data can be reached by increasing the $^{13}$C-pocket mass to $8 \times 10^{-3}$ $M_\odot$ in the D2 case, the corresponding match with the Ni isotope data becomes even worse. In principle, a consistent explanation of the Sr-Ba and Ni isotope data by the models can be found by further decreasing the $^{13}$C mass fraction while simultaneously increasing the $^{13}$C-pocket mass, based on the facts that (1) a Flat $^{13}$C-pocket of $8 \times 10^{-3}$ $M_\odot$ in the D2 case yields model predictions similar to a Flat $^{13}$C-pocket of $2 \times 10^{-3}$ $M_\odot$ in the ST case in Fig. 4a, and (2) the former case provides better match to the Ni isotope data than the latter case (Figs. 4b,c). However, $8 \times 10^{-3}$ $M_\odot$ is the maximum extension of the He-intershell predicted by FRUITY stellar model calculations for low-mass AGB stars. We thus could not find reasonable parameters for a Flat $^{13}$C-pocket model to simultaneously explain the Sr-Ba and Ni isotope data.

---

[6] Note that as will be discussed in Section 4.4, updated Torino model predictions for a 3 $M_{\odot}$, 1.5 $Z_{\odot}$ AGB star are almost identical to those for a 2 $M_{\odot}$, 1.5 $Z_{\odot}$ AGB star. Thus, the difference between the constrained amounts of $^{13}$C discussed here is mainly caused by the different stellar metallicities, i.e., 0.5 $Z_{\odot}$ versus 1.5 $Z_{\odot}$.

On the other hand, we observed that the Exponential pockets produce slightly steeper slopes in the Ni isotope plots and lower enrichments in neutron-rich isotopes (Fig. 5). This observation confirms our previous conclusion using the old FRANEC stellar models that the Flat $^{13}$C profile generally yields more variable model predictions than the Exponential $^{13}$C profile (Liu et al. 2014a, 2014b, 2015). The model predictions for Sr-Ba isotope ratios in the ST case (Fig. 5a) still follow the same trajectory with increasing $^{13}$C-pocket mass as the corresponding predictions with the Flat $^{13}$C-pockets, thus also providing good match to the grain data if the $^{13}$C-pocket mass is set to $(2-4) \times 10^{-3}$ $M_\odot$. However, model predictions by adopting the Exponential $^{13}$C profile cannot explain the Ni isotope data in Figs. 5b and 5c either. Although the slopes in the Ni three-isotope plots can be reproduced by reducing the $^{13}$C mass fraction to those in D3−D6 cases, the predicted $^{86}$Sr, $^{136}$Ba, and $^{61,62,64}$Ni enrichments are too low to match the majority of the grain data in Fig. 5. The Exponential $^{13}$C profile, therefore, also fails to explain all the grain data in a consistent way. Thus, we conclude that the inconsistency observed by Trappitsch et al. (2018) is not a result of the updated stellar models.

Nucci & Busso (2014) investigated magnetohydrodynamic processes at the base of convective envelopes using 2D and 3D analytical models. Trippella et al. (2016) then applied the developed formalism to investigate transport of H-envelope material to the He-intershell as a result of magnetic buoyancy, which subsequently induces the formation of the $^{13}$C-pocket. Based on "first principles," Trippella et al. (2016) were able to derive the formed $^{13}$C-pocket analytically (solid maroon line in Fig. 2), and the resultant Trippella $^{13}$C profile is quite distinctive with respect to the Exponential and Flat $^{13}$C profiles (Fig. 2). We therefore implemented the Trippella pocket in the updated Torino postprocessing AGB models to expand our parameter space for the $^{13}$C-pocket. Astonishingly, Fig. 6 shows that the inconsistency in the data-model comparisons for Sr-Ba and Ni isotopes observed earlier (Figs. 4 and 5) is well resolved by the same updated Torino model if the parameters for the Trippella pocket are adopted. Model predictions in the Original case are well overlapped with the center of the grain-concentrated region (i.e., the purple box) in Fig. 6a, and, in the meanwhile, closely reproduce the slopes of the Ni isotope data in Figs. 6b and 6c.

We further tested the effects of variable mass extensions and $^{13}$C mass fractions for the Trippella $^{13}$C-pocket based on the consideration that the derivation of the magnetic-buoyancy-induced $^{13}$C-pocket is based on the chosen magnetic field (and thus buoyancy velocity) and dynamic viscosity (Palmerini et al. 2018), and that variations in both parameters can lead to different amounts of $^{13}$C formed in the $^{13}$C-pocket. Note that we do not argue against the robustness of the results of Nucci & Busso (2014) by considering these variations, as the relative $^{13}$C profile within the pocket, which was derived by those authors from detailed three-dimensional magnetohydrodynamic simulations, indeed provides better match to the Sr-Ba and Ni isotope data than the Exponential and Flat $^{13}$C profiles and thus remains the same in these tests. Figure 6a clearly shows that the effect of varying the pocket mass on model predictions for Sr-Ba isotopes is a strong function of the $^{13}$C mass fraction, and reaches its minimum in the Original case. By slightly reducing the pocket mass by 33%, the data-model agreement can be further improved in Figs. 6b and 6c, given the consistency with the grain data in both the range of Ni isotope ratios and the corresponding slopes. Although we cannot exclude the possibility of variable pocket masses, a Trippella $^{13}$C-pocket with a pocket mass of $\leq 1.7 \times 10^{-3}$ $M_\odot$ seems unable to explain the isotope ratios of all the grains. For instance, with respect to a Trippella pocket of $5.0 \times 10^{-3}$ $M_\odot$ in the U1.5 case, Torino model calculations with a Trippella pocket of $1.7 \times 10^{-3}$ $M_\odot$ in the U1.5 case predict lowered $\delta^{88}$Sr and $\delta^{138}$Ba values and thus agree better with the grain data in Fig. 6a. The corresponding predictions for Ni isotope ratios, however, cannot cover the full range of the grain data in Figs. 6b and 6c. To conclude, the Sr-Ba and Ni isotope data together suggest that large Trippella pockets existed in the parent AGB stars of the majority of MS SiC grains, thus corresponding to efficient mixing of H into the He-intershell at low concentrations. For reference, the amount of $^{13}$C contained in a Trippella pocket of $3.3 \times 10^{-3}$ $M_\odot$ in the Original case corresponds to $6.74 \times 10^{-6}$ $M_\odot$.

In addition, Fig. 6 further suggests that variation in the amount of $^{13}$C within the Trippella $^{13}$C-pocket is limited, because model predictions in the Original case consistently match the majority of the MS grain data for their Sr-Ba and Ni isotope ratios. More importantly, model predictions with varying pocket masses in the Original case further help to explain the distribution of the grain data in Fig. 6a. Thus, if the pocket mass indeed varied by a factor of three in the parent AGB stars of MS SiC grains, the predicted variability in the slopes of Ni isotope ratios would well account for the amount of scatter from the lines defined by the grains (gray bands in Figs. 6b and 6c). Consequently, this would imply that the $^{13}$C mass fractions are quite comparable across the parent AGB stars of MS SiC grains. It is also noteworthy to point out that the majority of the Ni isotope data in Fig. 6 clearly define one single $s$-process end-member with limited variability, which is also supported by the fact that the majority of Sr-Ba data are clustered within the purple box despite the large analytical errors. In comparison, model calculations by adopting variable $^{13}$C-pockets, especially variable $^{13}$C mass fractions, predict a wide range of $s$-process end-members. Therefore, previous arguments that heavy-element isotopic data of MS SiC grains require a wide range of $^{13}$C mass fractions (e.g., Lugaro et al. 2003) and/or large variations in the mass loss rate (Palmerini et al. 2018) do not seem to hold true any more, especially because the high-precision Ni isotope data fall along straight lines with limited scatter in Figs. 6b and 6c.

## 4. DISCUSSION

### 4.1. Effects of GCE and Ni Contamination

GCE is an evolutionary process of the chemical composition of the Galaxy, resulting from newly synthesized nuclei contributed by successive generations of stars to the interstellar medium. The starting composition of an AGB star of a given metallicity is therefore determined by the GCE effect as a function of time and/or Galactic location. Alexander & Nittler (1999) previously observed that MS grains show linearly correlated Si and Ti (mainly light Ti isotopes, e.g., $^{46}$Ti/$^{48}$Ti) isotopic compositions, most likely reflecting the average GCE trends of these elements. Trappitsch et al. (2018) discussed the GCE effect on $\delta^{60}$Ni and concluded that the variations in the $\delta^{60}$Ni values of MS SiC grains are mainly affected by GCE rather than the $s$-process in AGB stars. This is because the AGB models predict only up to 50‰ variability in $\delta^{60}Ni$ as a result of the *s*-process nucleosynthesis (Fig. 7a), while the grains show much larger variations (~200‰) that are roughly correlated with their $\delta^{29}Si$ values as a result of GCE, especially if Ni contamination is taken into account for the close-to-solar Ni isotope ratios observed in a few grains (Trappitsch et al. 2018). In contrast, the $\delta^{61}Ni$ and $\delta^{62}Ni$ data do not show any correlation with the $\delta^{29}Si$ data (e.g., Fig. 7b), but instead are well correlated with the $\delta^{64}Ni$ data (e.g, Figs. 6b,c) with limited scatter. This indicates that the GCE effects on the $\delta^{61}Ni$, $\delta^{62}Ni$, and $\delta^{64}Ni$ data for MS grains, if present, are negligible.

On the other hand, it is noteworthy to point out that Ni contamination might have affected some of the Ni isotope grain data in Fig. 7. However, since Trappitsch et al. (2018) observed a gross correlation between the $\delta^{60}Ni$ and $\delta^{29}Si$ values, it is clear that Ni contamination, if indeed sampled during the analysis, did not significantly affect the collected Ni isotope data. Also, the effect of Ni contamination with solar system Ni on a Ni three-isotope plot is to pull a data point toward the solar composition, i.e., the crossover of the dashed lines, in Figs. 6b and 6c, which therefore does not affect the slopes in the plots. Thus, although we cannot completely exclude the possibility of Ni contamination, which might have slightly lowered the most extreme Ni isotope anomalies toward the solar composition, such contamination should not have affected the Ni isotope slopes defined by all the grains (e.g., Figs. 6b,c). Consequently, this implies that although the possibility of Ni contamination precludes constraining the maximum $^{13}C$-pocket mass using the Ni isotope data, it does not affect our conclusion that the Trippella $^{13}C$ profile explains simultaneously the Sr-Ba data and the Ni isotope slopes better than the other considered $^{13}C$-pocket profiles.

### 4.2. Effect of Stellar Metallicity

We explored the effect of metallicity on our results by varying the metallicity for 3 $M_\odot$ AGB stars with different Trippella pocket cases (Fig. 8). Although model predictions for a 1.5 $Z_\odot$ AGB star can only consistently explain the Sr-Ba and Ni isotope data in the Original case, the model predictions remain almost the same for 0.5 and 1.0 $Z_\odot$ AGB stars if the $^{13}C$ mass fraction is lowered by a factor of 1.5 and 2, respectively. This means that the $^{13}C$ mass fraction that best agrees with the grain data is a linear function of the chosen stellar metallicity, and vice versa. Unfortunately, the $^{13}C$ mass fraction cannot be estimated robustly, because the value, i.e., $10^{-5}$ $M_\odot$ $^{13}C$ contained within a Trippella pocket of $5 \times 10^{-3}$ $M_\odot$, reported by Trippella et al. (2016) was derived based on a number of chosen parameters and assumptions, e.g., the magnetic field strength in the He-intershell, which are not well constrained by astronomical observations. The Sr-Ba and Ni isotopic compositions of MS SiC grains, therefore, cannot be directly used to constrain the stellar metallicities of their parent AGB stars.

This conclusion in fact is in contrast to a recent speculation that MS SiC grains could have originated from supersolar-metallicity AGB stars based on data-model comparison for Sr, Zr, and Ba isotope ratios (Lugaro et al. 2018). The contradiction mainly results from different treatments of the $^{13}C$-pocket adopted in the nucleosynthesis calculations between the two studies. While here we treated the pocket mass, the $^{13}C$ mass fraction, and the $^{13}C$ profile all as free parameters in our calculations given the aforementioned reasons, Lugaro et al. (2018) adopted an Exponential $^{13}C$-pocket with fixed $^{13}C$ mass fractions across the $^{13}C$-pocket and tested the effect of varying $^{13}C$-pockets by changing the pocket mass only. As explained earlier, the model predictions are essentially a linear function of the neutron-to-seed ratio, and the $^{13}C$ mass fraction plays a significant role in determining the *s*-process isotopic signatures. Thus, one cannot simply derive a constraint on the stellar metallicity without first constraining a priori $^{13}C$ mass fraction chosen for study. This point is in fact well illustrated by our calculations shown in Figs. 6 and 8 that 3 $M_\odot$ model predictions with metallicities ranging from 0.5 $Z_\odot$ to 1.5 $Z_\odot$ can all provide excellent match to the grain data if proper $^{13}C$ mass fractions are chosen. On the other hand, sophisticated magnetohydrodynamic simulations of the $^{13}C$-pocket formation, which consider all possible physical mechanisms, are needed in the future to pin down the $^{13}C$ mass fraction. If such simulations can constrain the $^{13}C$ mass fraction to lie above the D1.5 case in low-mass AGB stars with the assist of observational constraints on adopted physical parameters (e.g., rotational velocity, magnetic field), this information can indeed be directly translated to the fact that MS SiC grains came from low-mass AGB stars with supersolar metallicities, which in the meantime provides an explanation to the supersolar Si isotope ratios of MS SiC grains. It is noteworthy to mention that although Lugaro et al. (2018) concluded that their 2 $Z_\odot$ models provide a better match to the Sr, Zr, and Ba grain data than the corresponding 1 $Z_\odot$ models, Lewis et al. (2013) in fact previously pointed out that the expected frequency for MS grains coming from 2 $Z_\odot$ parent stars is below 5%, according to the metallicity-age relation derived for MS grains using their Si isotope ratios.

Finally, we would like to highlight the point that MS SiC grains should have originated from low-mass AGB stars with quite similar metallicities, because the grain data point to limited variations in the neutron-to-seed ratios in their parent stars. In turn, this requires limited variabilities of both the $^{13}C$ mass fraction and the stellar metallicity, mainly because to our knowledge, no physical mechanism has been identified that could vary the $^{13}C$ mass fraction as a linear function of the stellar metallicity. This implication in fact is consistent with the conclusion of Lewis et al. (2013) that the majority of SiC grains should have formed in AGB stars with metallicities close to (or slightly above) the solar value.

### 4.3. Effects of Nuclear Uncertainties

We also computed the varying-metallicity models shown in Fig. 8 using the K94 rate (larger colored symbols) to evaluate its effect on our constrained $^{13}C$-pocket parameters. This is because the recently recommended $^{22}Ne(\alpha,n)^{25}Mg$ rates by nuclear experiments lie between ½ × K94 and K94 rates at AGB stellar temperatures (Jaeger et al. 2001; Longland 2012). Comparison with the corresponding ½ × K94 calculations shows that the K94 rate leads to more efficient operation of the $^{22}Ne(\alpha,n)^{25}Mg$ reaction, thus increasing the model predictions for $\delta^{88}Sr$ because of reduced $^{86}Sr$ production and simultaneously producing steeper slopes in Figs. 8b and 8c as a result of the branch point at $^{60}Co$. Overall, the K94 rate further improves the match of the model predictions with the grain data and therefore has no effect on our derived $^{13}C$-pocket constraints. Note that it was shown previously that FRUITY model calculations predict steeper slopes in the Ni three-isotope plots than do Torino model calculations (Trappitsch et al. 2018). This is because FRUITY stellar models predict that a portion of $^{13}C$ remains unburned during the interpulse and is engulfed into the convective zone generated by the TP, while Torino model calculations assume that $^{13}C$ is fully consumed during the interpulse. In turn, the leftover $^{13}C$ in the FRUITY models is subsequently burned at higher temperature during the next TP and thus releases neutrons at a higher density, which more efficiently activates branching points and therefore produces steeper slopes in the Ni three-isotope plots. In addition, the match given by the model predictions with the grain data in Fig. 8 can be further improved by slightly lowering the $^{13}C$ mass fraction (green squares with lines in Fig. 9) and by increasing the $^{64}Ni$ MACS value by 10% (blue circles with lines in Fig. 9) given its uncertainty reported in KADoNiS v0.3. The differences between the two models in the Ni three-isotope plots in Fig. 9, however, are quite small, and we therefore still adopt the recommended $^{64}Ni$ MACS value in the following discussion.

### 4.4. Effect of Stellar Mass

Observations of AGB stars and planetary nebulae constrain the masses of C-rich stars to lie between ∼1.5 and ∼3−4 $M_\odot$ in the Magellanic Clouds (Frogel et al. 1990; Lattanzio & Wood 2003). For reference, the FRUITY stellar code predicts that 1.5 $M_\odot$ AGB stars become C-rich in the last several TPs only at ≤0.5 $Z_\odot$ and that AGB stars become C-rich for stellar masses up to 3 $M_\odot$. The FRUITY models are quite similar to the Monash models recently used by Lugaro et al. (2018) up to 3 $M_\odot$. A clear discrepancy then occurs above 3 $M_\odot$. While the Monash code predicts C-rich TPs for 3−4 $M_\odot$ AGB stars, the FRUITY stellar code predicts that 4 $M_\odot$ stars are always O-rich during the AGB phase. The discrepancy is likely caused by differences in the treatment for the radiative/convective interface at the base of the convective envelope as well as in the adopted mass-loss law. However, a definitive answer requires a systematic comparison of the two stellar codes under scrutiny, which is beyond the scope of this paper. Instead, we investigate the effect of varying the stellar mass in the range of 1.5−3 $M_\odot$ using the 0.5 $Z_\odot$ FRUITY stellar models.

Figure 10 clearly illustrates that variation in the stellar mass (between 1.5 $M_\odot$ and 3 $M_\odot$) has a negligible effect on our constrained [13]C-pocket parameters, especially given the almost identical model predictions for 2 $M_\odot$ and 3 $M_\odot$ AGB stars. The observed similarity between 2 $M_\odot$ and 3 $M_\odot$ FRUITY AGB model predictions is supported by the results of Lugaro et al. (2018) that large variations in the model predictions are only observed when the stellar mass lies above 3 $M_\odot$. This is likely because the Monash models for >3 $M_\odot$ AGB stars predict much higher stellar temperatures, corresponding to more efficient operation of the minor neutron source, $^{22}Ne(\alpha,n)^{25}Mg$ reaction. As a result, the branch points at $^{85}Kr$ and $^{136}Cs$ become activated, resulting in increased model predictions for $\delta^{88}Sr$ and $\delta^{137}Ba$, and decreased predictions for $\delta^{134}Ba$ (Liu et al. 2014a, 2015). It is noteworthy to point out that although the Monash model predictions presented by Lugaro et al. (2018) provide a fairly good match to the grain data for their Zr isotope ratios, in fact none of the model predictions provides a satisfactory match to the center of the grain concentrated region in the Sr-Ba isotope plot. This points to the different sensitivities of these isotope ratios to the different parameters in the AGB stellar models. As explained earlier, $^{88}Sr$ and $^{138}Ba$ are the bottlenecks along the s-process path, while the Ni isotopes sit next to the Fe seeds for neutron capture during the s-process. As a result, all of the Sr-Ba and Ni isotope ratios are more sensitive to the [13]C-pocket parameters adopted in the stellar model, and the Trippella [13]C profile seems to be the best solution to consistently explain the Sr-Ba and Ni isotope data, according to the updated Torino model calculations. Finally, the poor agreement with the grain data by the 1.5 $M_\odot$ AGB model predictions in Fig. 10 mainly results from the reduced number of C-rich TPs as well as the lowered stellar temperature in the last C-rich TP. Since 1.5 $M_\odot$ AGB model predictions alone cannot explain all the MS SiC grain data, it is highly likely that most of the grains formed in >1.5 $M_\odot$ AGB stars.

### 4.5. Constraints from Other Isotope Ratios

We further verified the constrained [13]C-pocket parameters by comparing the corresponding model predictions with the grain data for additional Ba, and all Zr and Ti isotope ratios in Figs. 11−13. Given the large uncertainties in measured $\delta^{134}Ba$ values due to the low abundance of $^{134}Ba$, it is impossible to evaluate the quality of the data-model match in Fig. 11a. On the other hand, the model predictions with the constrained [13]C-pocket parameters are in reasonably good agreement in Fig. 11b. Finally, the updated Torino model predictions for Zr isotope ratios in Fig. 12 agree well with the model predictions for 2 $Z_\odot$ AGB stars with 2−3 $M_\odot$ in Fig. 2 of Lugaro et al. (2018). However, while Lugaro et al. (2018) explained the grains with closer-to-solar Zr isotope ratios using model predictions for higher-mass AGB stars with more efficient operation of the $^{22}Ne(\alpha,n)^{25}Mg$ reaction (i.e., more positive $\delta^{90,91,92,96}$Zr values), our model calculations suggest that the grains with closer-to-solar Zr isotope ratios could be alternatively explained by smaller $^{13}$C-pockets ($<1.7\times10^{-3}$ $M_\odot$). On the other hand, although Fig. 12 shows that the variability of the $^{13}$C-pocket constrained by the Sr-Ba and Ni isotope data cannot explain the range of the Zr isotope data, this figure also clearly points out significant inconsistencies in $\delta^{90}$Zr and $\delta^{91}$Zr values between the two sets of grain data in the literature (Nicolussi et al. 1997; Barzyk et al. 2007). The Nicolussi et al. data are generally more positive and less scattered than the Barzyk et al. data, and are in fact more consistent with the model predictions. Given the inconsistency in the literature data, Zr isotope data of higher quality and higher precision are urgently needed to better distinguish between the effects of the stellar mass and the $^{13}$C-pocket mass. Finally, $^{50}$Ti also has a magic number (28) of neutrons and thus extremely small MACS values. The model predictions with the constrained $^{13}$C-pocket parameters explain the Ti isotope data of MS grains[7] (Fig. 13). The fact that a few grains deviate from the linear fit in Fig. 13b indicates that in addition to the *s*-process nucleosynthesis, the $\delta^{49}$Ti values of AGB stars are probably also affected by GCE but to a lesser extent. In comparison, the grain data are well correlated in Fig. 13a, and the same model predictions fail to explain the full range of the grain data in this figure. Previous studies showed that the $\delta^{46}$Ti and $\delta^{47}$Ti values of MS grains are well correlated with their $\delta^{29}$Si values, which indicates that the abundances of $^{46}$Ti and $^{47}$Ti in AGB stars are dominantly affected by GCE (Alexander & Nittler 1999; Gyngard et al. 2018; Nguyen et al. 2018). This agrees with the fact that the predicted *s*-process effects are less than 50‰ in Fig. 13a while the grains exhibit much larger variations.

## 5. CONCLUSION

We have compared literature Ti, Ni, Sr, Zr, and Ba isotope ratios of presolar MS SiC grains from AGB stars with updated Torino postprocessing nucleosynthesis model predictions based on full evolutionary FRUITY AGB stellar models. We found that the large Flat and Exponential $^{13}$C-pockets that successfully matched the Sr-Ba isotope data in the study of Liu et al. (2015) fail to consistently explain the Ni isotope data of MS SiC grains from Trappitsch et al. (2018). However, we found that this inconsistency can be well reconciled by adopting the magnetic-buoyancy-induced $^{13}$C-pocket proposed by Trippella et al. (2016). The grain data thus suggest deep mixing of H into the He-intershell from the envelope at low concentration(s) occurring in the parent AGB stars of MS SiC grains. We also showed that the agreement between the grain data and the model by adopting the Trippella pocket is unaffected by uncertainties in the stellar mass and the $^{22}$Ne($\alpha,n$)$^{25}$Mg rate. On the other hand, we observed that the updated Torino model predictions for low-mass C-rich AGB stars depend on the neutron-to-seed ratio. This means that when the Trippella pocket is adopted, the grain data can be equally well explained by Torino model calculations for AGB stars with a wide range of stellar metallicities if proper $^{13}$C mass fractions are chosen. Thus, we want to emphasize here that the heavy-element isotopic compositions of MS SiC grains cannot provide direct constraints on the stellar metallicities of their parent AGB stars because of the coupled effect of the stellar metallicity and the amount of $^{13}$C contained in the $^{13}$C-pocket on AGB model predictions for the *s*-process nucleosynthesis.

Finally, we highlight four directions here for future studies of *s*-process nucleosynthesis to further test the validity of the constrained $^{13}$C-pockets here: (1) New high-precision Zr isotope data in MS SiC grains are urgently needed to better distinguish between the effects of the stellar mass and the pocket mass, which will allow us to put an upper limit on the stellar mass of parent AGB stars of MS SiC grains. (2) The effect of the pocket mass on Torino model calculations can be further tested by considering the shrinkage of the $^{13}$C-pocket with increasing TPs during the evolution of an AGB star as predicted by the FRUITY models (Cristallo et al. 2009). (3) The effect of uncertainties in AGB stellar models need to be tested in the future by implementing the constrained $^{13}$C-pockets from this study in other AGB stellar codes, e.g, the Monash (Karakas & Lattanzio 2014) and NuGrid (Pignatari et al. 2016) codes. (4) The validity of the conclusion from this study based on postprocessing nucleosynthesis calculations can be further tested in the future by running stellar evolutionary models coupled to a full nucleosynthesis network (as FRUITY models) as well as by implementing a detailed treatment of the mixing induced by magnetic buoyancy in the stellar models.

---

[7]Given the large number of MS grains with high precision Ti isotope data reported in Gyngard et al. (2018), we chose this set of data for the data-model comparison. A comparison of literature data by Gyngard et al. (2018) shows general agreements between different datasets.

Acknowledgements: We thank the anonymous reviewer for constructive review that helped improve the paper. This work was supported by NASA (grants NNX17AE28G to LRN and 80NSSC17K0251 to AMD). Part of this work was performed under the auspices of the U.S. Department of Energy by Lawrence Livermore National Laboratory under Contract DE-AC52-07NA27344 and by the Laboratory Directed Research and Development Program at LLNL under project 17-ERD-001. LLNL-JRNL-750602

## Figure Caption

**Fig. 1**. Position (in mass coordinate) versus time (in the unit of years) of (1) the inner border of the convective envelope (black line), (2) the location of the maximum energy production within the H-burning shell (red line), (3) the inner border of the He-intershell (magenta broken line), and (4) the maximum energy production within the He-burning shell (blue line) of a 3 $M_\odot$, 1.5 $Z_\odot$ AGB star given by FRUITY model calculations. The blue plateaus between the numbered thermal pulses correspond to activation of the $^{13}C(\alpha,n)^{16}O$ reaction (see text for more detail).

**Fig. 2**. The $^{13}C$ mass fraction ($X(^{13}C)$) is plotted against the stellar mass in units of $M_\odot$ proceeding from the bottom of the H envelope (right) to the top of the He-intershell (left) of an AGB star. The stellar mass coordinate at the interface of the two layers is defined as zero. Different $^{13}C$ profiles are explained in the text and shown as different colored lines. A smaller Trippella pocket ($3.3 \times 10^{-3}$ $M_\odot$) in the D3 case, which might arise from a weaker magnetic field or interactions between magnetic and hydrodynamic effects, is shown in (b) to illustrate how the mass fraction and the mass extension are varied in Torino models to test their effects on $s$-process model predictions. The three-zones of the Exponential $^{13}C$-pocket are labeled in yellow (Zone-I), green (Zone-II), and red (Zone-III) in (c) in logarithmic scale.

**Fig. 3**. Four-isotope plot of $\delta^{88}Sr$ vs. $\delta^{138}Ba$ values for presolar MS SiC grains from Liu et al. (2015) and Stephan et al. (2018). The 2D joint histogram of the data is shown using a yellow-to-blue color scheme to reflect the grain density, and the number next to the color scheme represents the corresponding number of grains. The grain-concentrated region is highlighted by a purple box.

**Fig. 4**. (a) Four-isotope plot of $\delta^{88}Sr$ vs. $\delta^{138}Ba$, and three-isotope plots of (b) $\delta^{61}Ni$ vs. $\delta^{64}Ni$ and (c) $\delta^{62}Ni$ vs. $\delta^{64}Ni$. The purple box represents the highest density of the MS grain data from Fig. 3. Updated Torino model predictions by adopting Flat $^{13}C$-pockets with varying $^{13}C$ mass fractions (D2−U1.3) and pocket masses ($1 \times 10^{-3}$–$8 \times 10^{-3}$ $M_\odot$) for a 3 $M_\odot$, 1.5 $Z_\odot$ AGB star are compared to 47 MS (10 of them are unclassified) SiC grain data from Liu et al. (2015), 3 MS SiC from Stephan et al. (2018), and 56 MS SiC grain data from Trappitsch et al. (2018). The entire evolution of the AGB envelope composition is shown as a line, but symbols for the pulses are plotted only when C > O in the envelope and condensation of SiC is possible. The ½ × K94 $^{22}Ne(\alpha,n)^{25}Mg$ rate is adopted in all model calculations (see text).

**Fig. 5**. The same set of grain data are compared to updated Torino model predictions by adopting Exponential $^{13}C$-pockets with varying pocket masses ($1 \times 10^{-3}$–$8 \times 10^{-3}$ $M_\odot$) in varying cases for a 3 $M_\odot$, 1.5 $Z_\odot$ AGB star.

**Fig. 6**. The same set of grain data are compared to updated Torino model predictions by adopting Trippella $^{13}C$-pockets with varying $^{13}C$ mass fractions (D3−U1.5) and pocket

masses ($1.7 \times 10^{-3}$–$5.0 \times 10^{-3}$ $M_\odot$) for a 3 $M_\odot$, 1.5 $Z_\odot$ AGB star. The best-fit lines with 95% confidence bands for the Ni isotope ratios are shown in figures (b) and (c), which are calculated using the weighted orthogonal distance regression (ODR) method. The weighted ODR method is a linear least-squares fitting method that minimizes scatter orthogonal to the best-fit line and considers uncertainties in both the x- and y-axis, for each data point.

**Fig. 7.** (a) Three-isotope plot of $\delta^{60}$Ni versus $\delta^{64}$Ni, and (b) four-isotope plot of $\delta^{60}$Ni versus $\delta^{29}$Si. The same set of the Torino model predictions in Fig. 6 are compared to the MS grain data from Trappitsch et al. (2018).

**Fig. 8.** The same set of grain data as in previous figures are compared to updated Torino model predictions by adopting $3.3 \times 10^{-3}$ $M_\odot$ Trippella $^{13}$C-pockets in varying cases for 3 $M_\odot$ AGB stars with 0.5 $Z_\odot$ and 1.0 $Z_\odot$.

**Fig. 9.** The same set of grain data as in previous figures are compared to updated Torino model predictions by adopting $3.3 \times 10^{-3}$ $M_\odot$ Trippella $^{13}$C-pockets in the D2 case for a 3 $M_\odot$, 1.0 $Z_\odot$ AGB star.

**Fig. 10.** The same set of grain data as in previous figures are compared to updated Torino model predictions by adopting the $3.3 \times 10^{-3}$ $M_\odot$ Trippella $^{13}$C-pocket in the D3 case for 0.5 $Z_\odot$ AGB stars with 1.5 $M_\odot$, 2.0 $M_\odot$, and 3.0 $M_\odot$.

**Fig. 11.** Three-isotope plots of (a) $\delta^{134}$Ba and (b) $\delta^{137}$Ba vs. $\delta^{135}$Ba. The updated Torino model predictions that best fit the Sr-Ba and Ni isotope data in Figs. 6 and 8 are compared to MS SiC grain data from Liu et al. (2014a, 2015) and Stephan et al. (2018). For the *r*-mostly isotope $^{135}$Ba, its initial abundance varies with the stellar metallicity because of consideration of GCE (Bisterzo et al. 2011).

**Fig. 12.** Three-isotope plots of (a) $\delta^{90}$Zr, (b) $\delta^{91}$Zr, and (c) $\delta^{92}$Zr vs. $\delta^{96}$Zr. The updated Torino model predictions for 1.0 $Z_\odot$ and 1.5 $Z_\odot$ AGB stars that best fit the Sr-Ba and Ni isotope data in Figs. 6 and 8 are compared to grain data from Nicolussi et al. (1997) and Barzyk et al. (2007).

**Fig. 13.** Three-isotope plots of (a) $\delta^{47}$Ti vs. $\delta^{46}$Ti, and (b) $\delta^{49}$Ti vs. $\delta^{50}$Ti. The updated Torino model predictions for 1.0 $Z_\odot$ AGB stars that best fit the Sr-Ba and Ni isotope data in Figs. 6 and 8 are compared to MS grain data from Gyngard et al. (2018) with 2σ errors plotted.

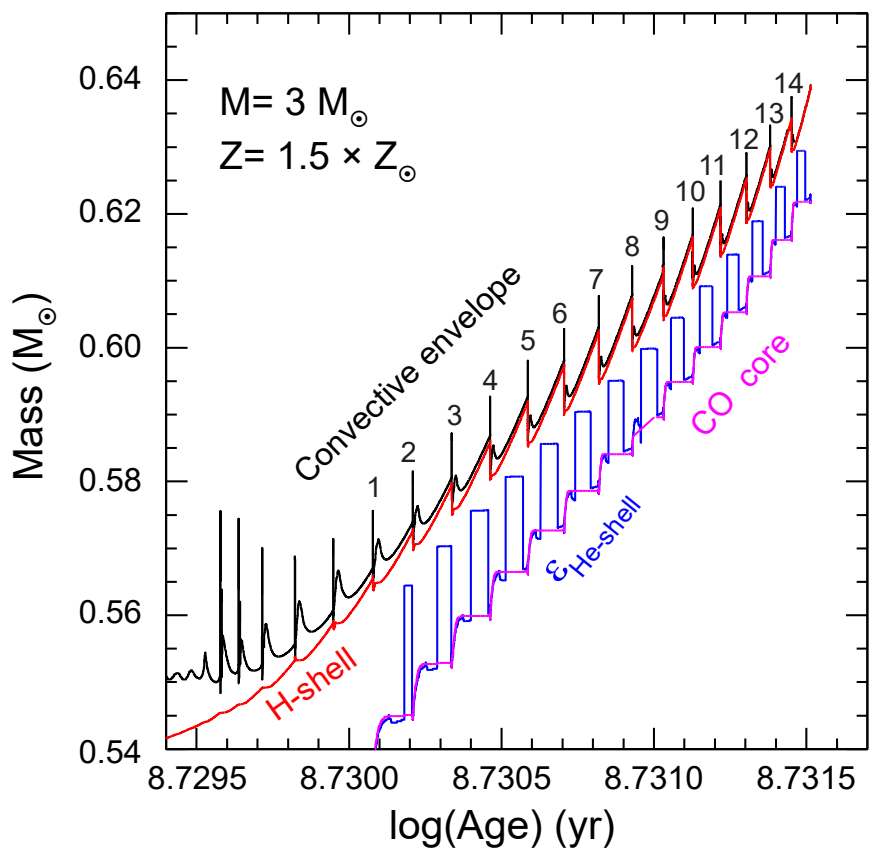

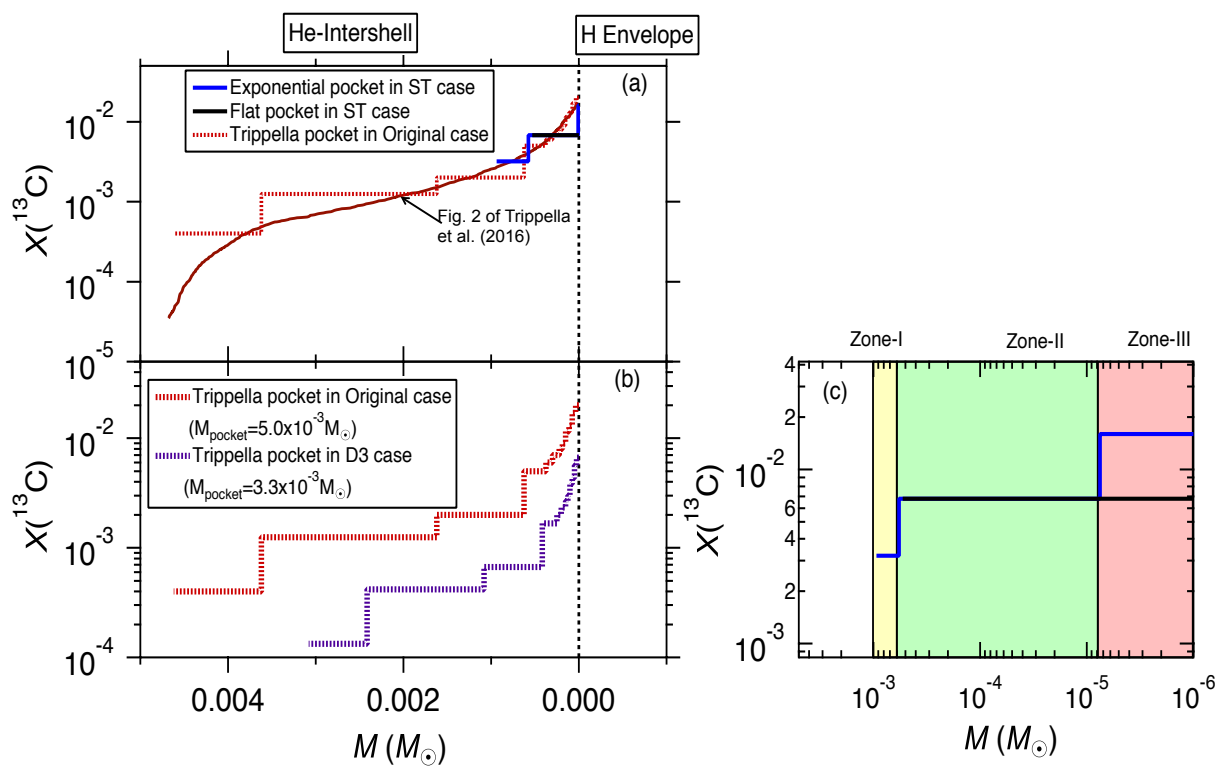

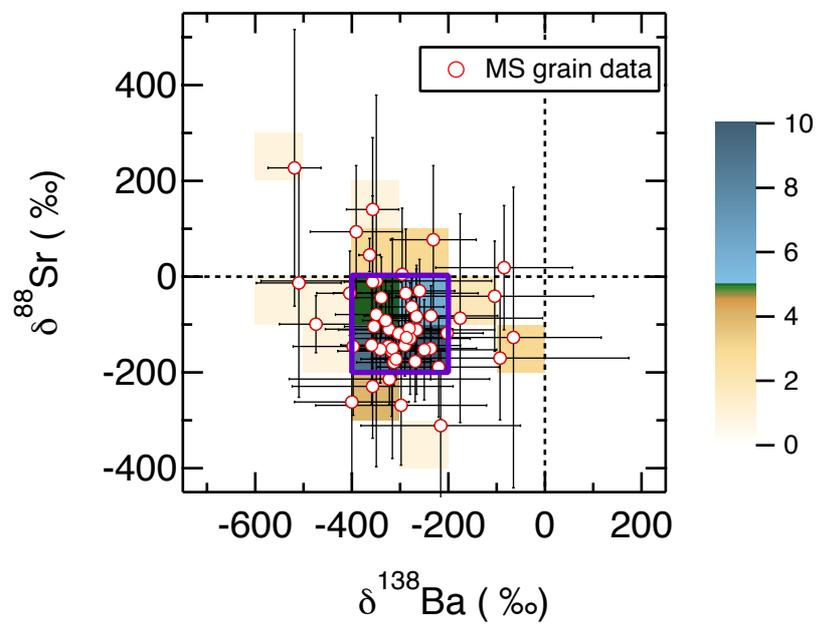

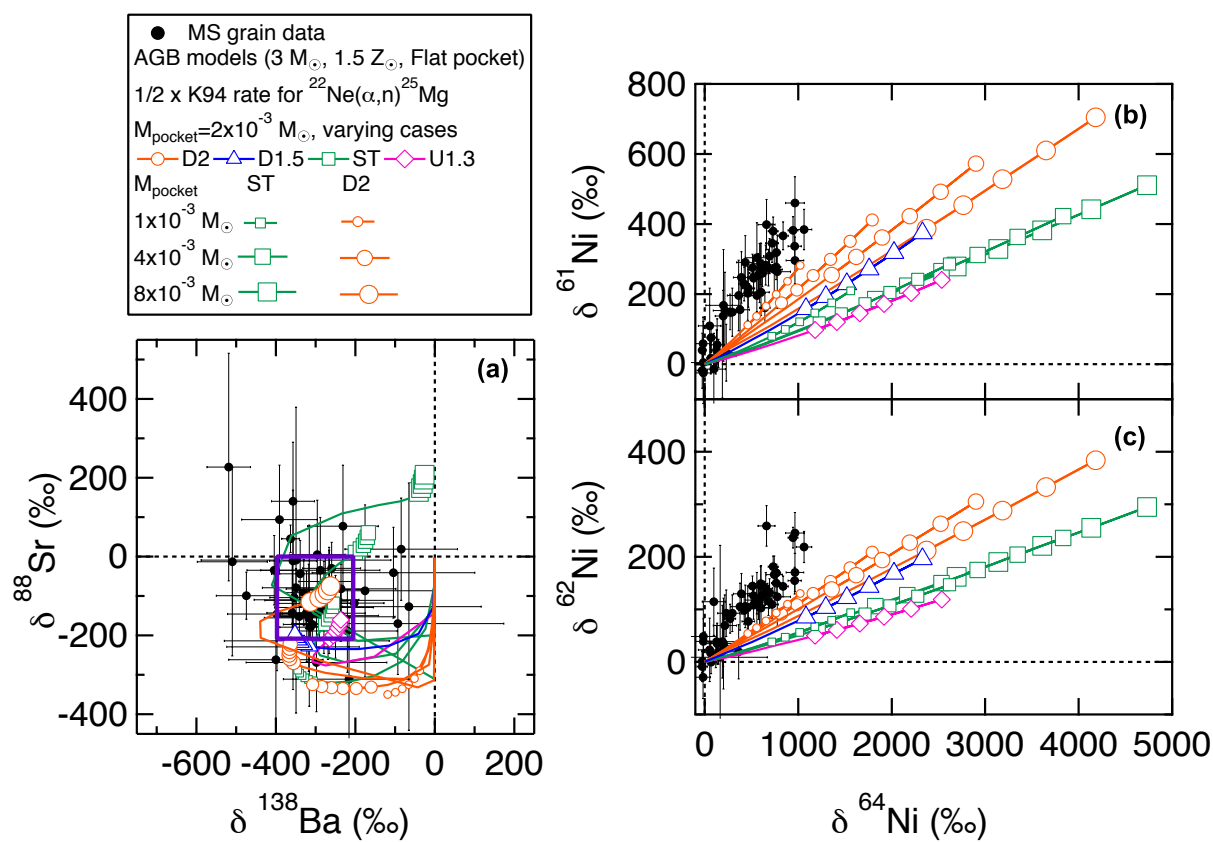

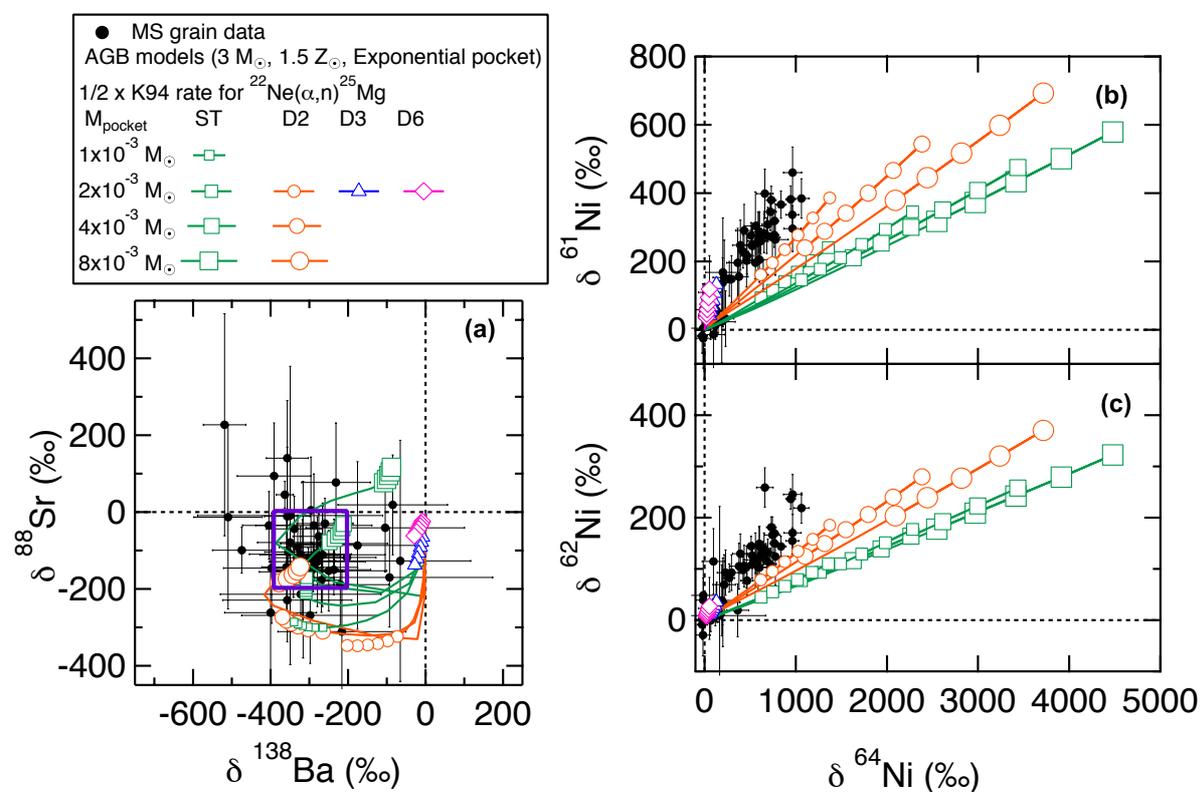

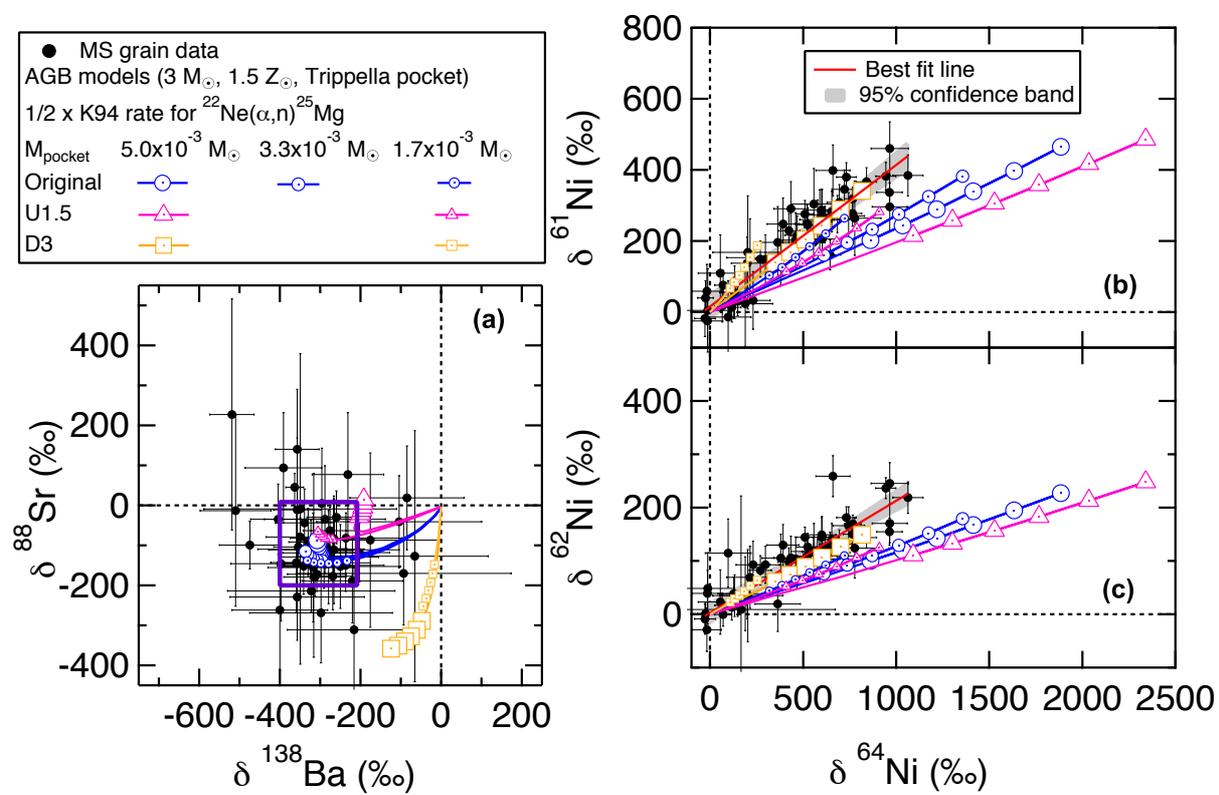

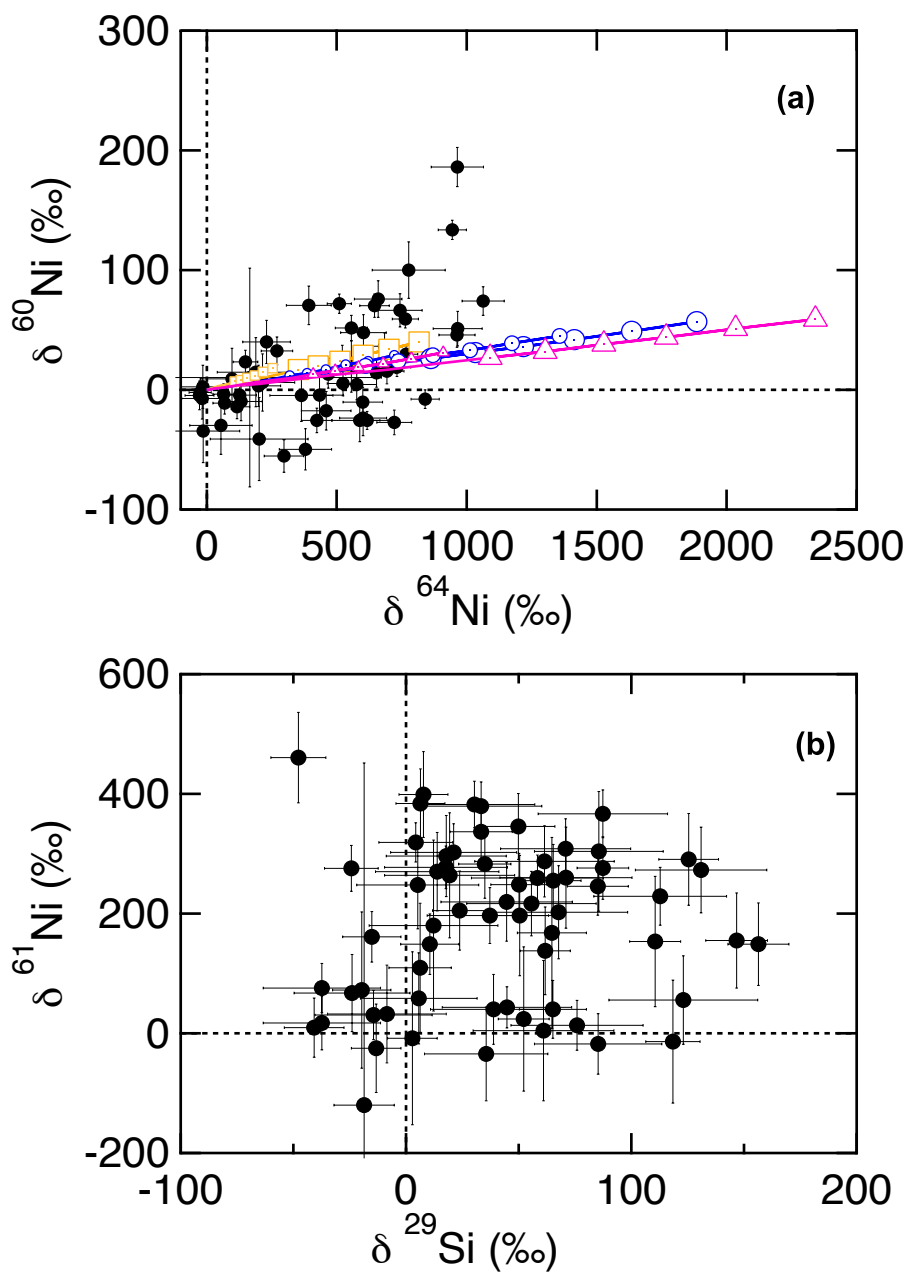

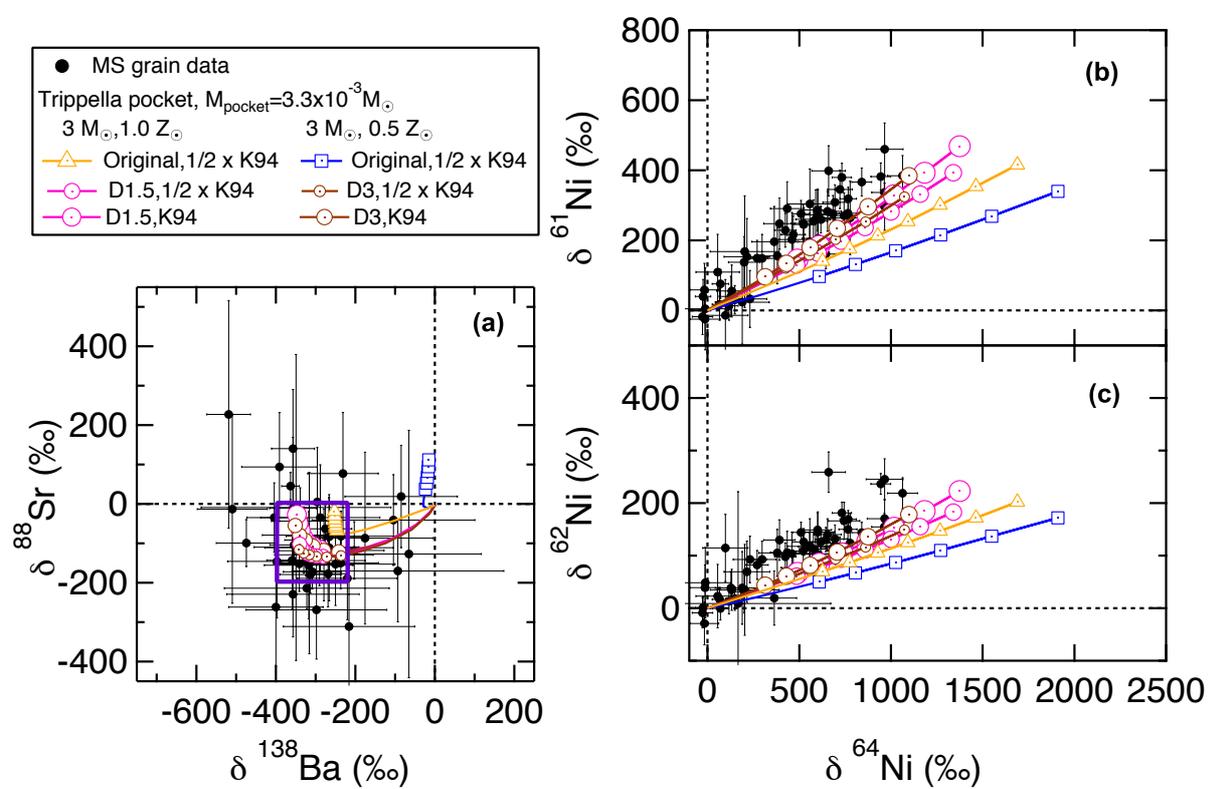

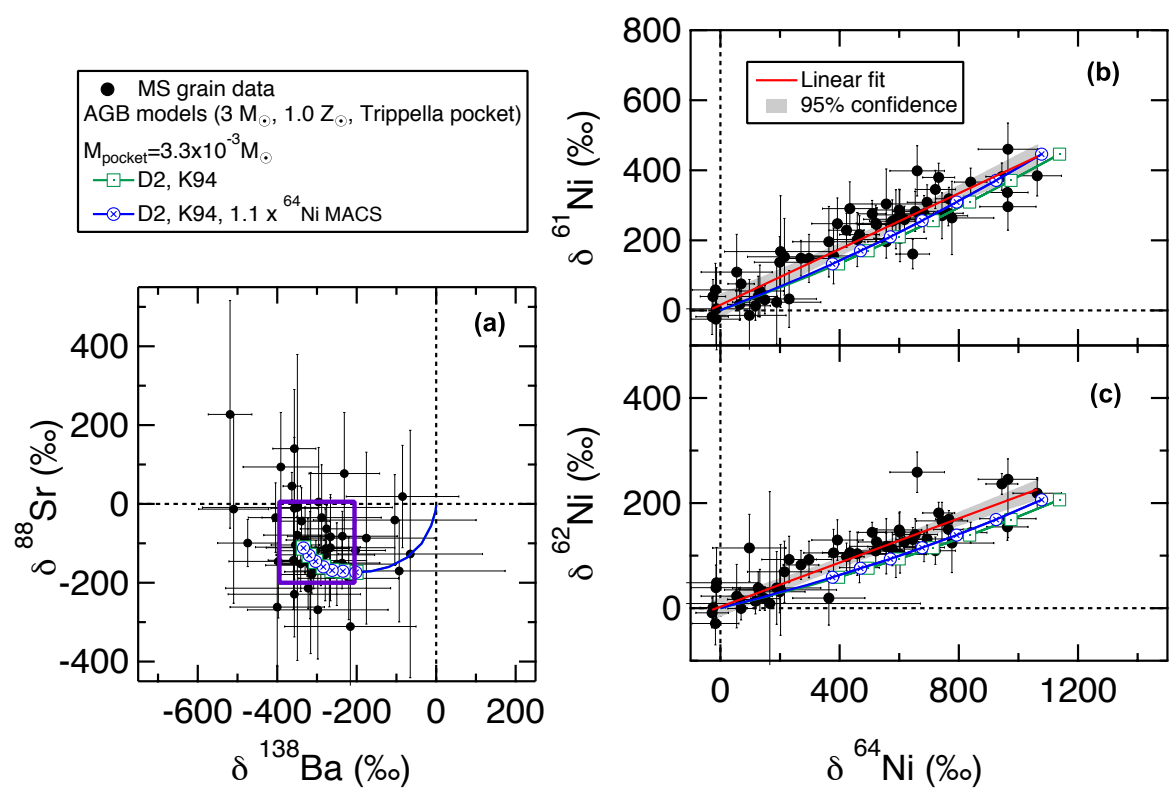

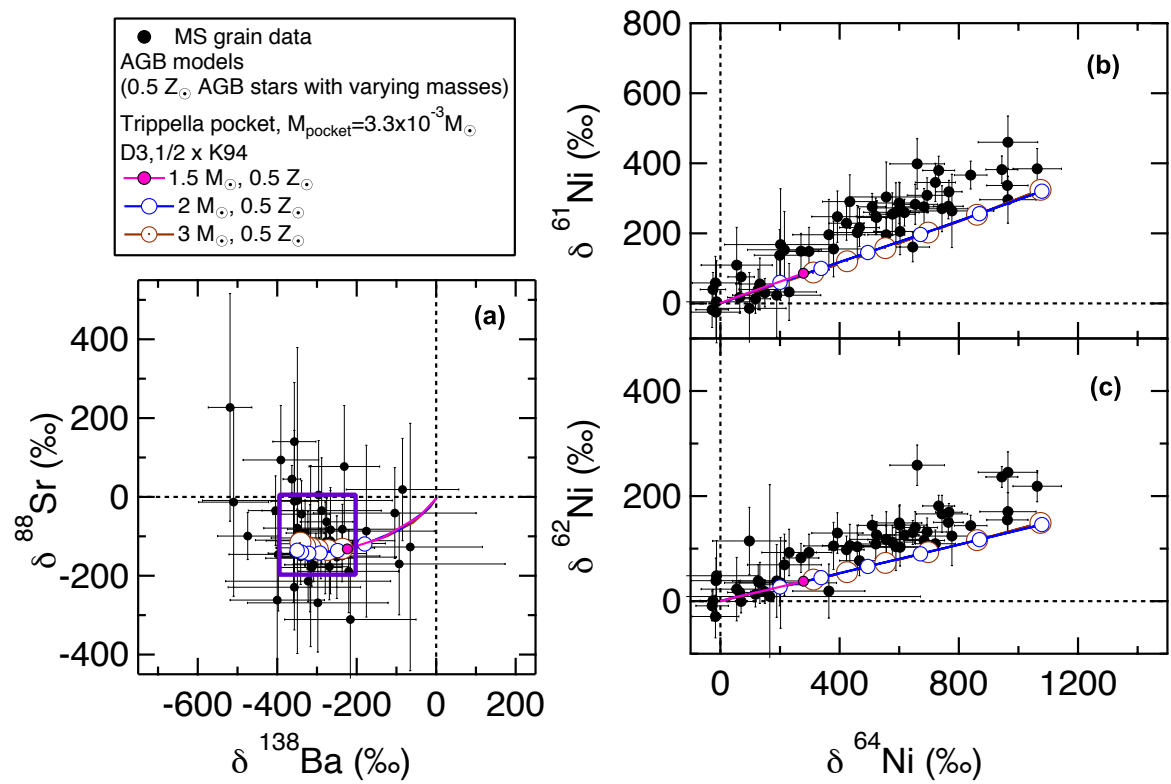

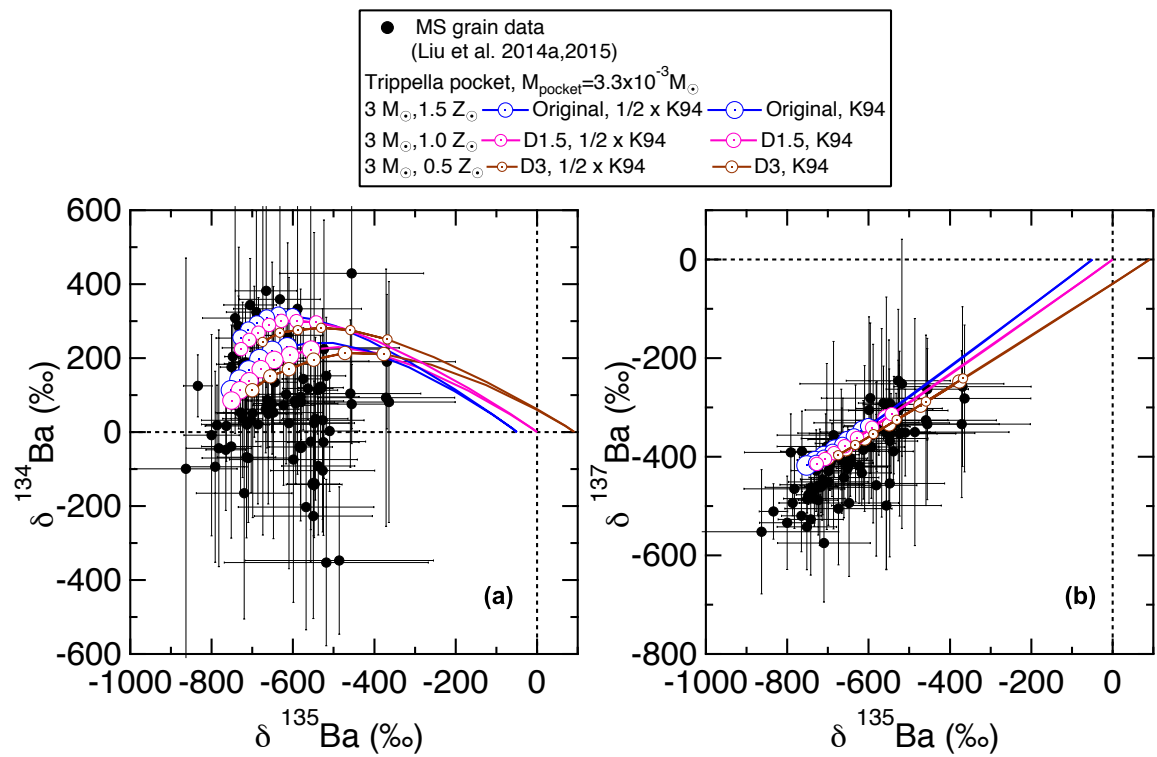

Fig. 12

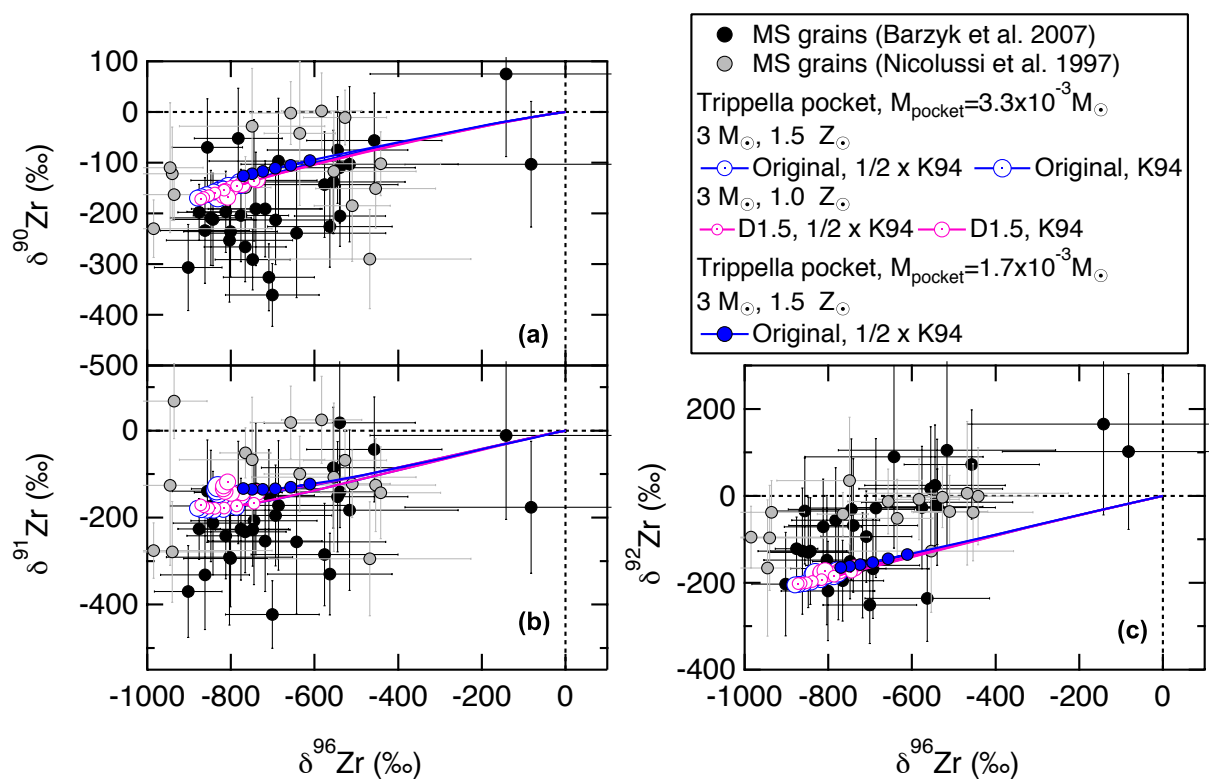

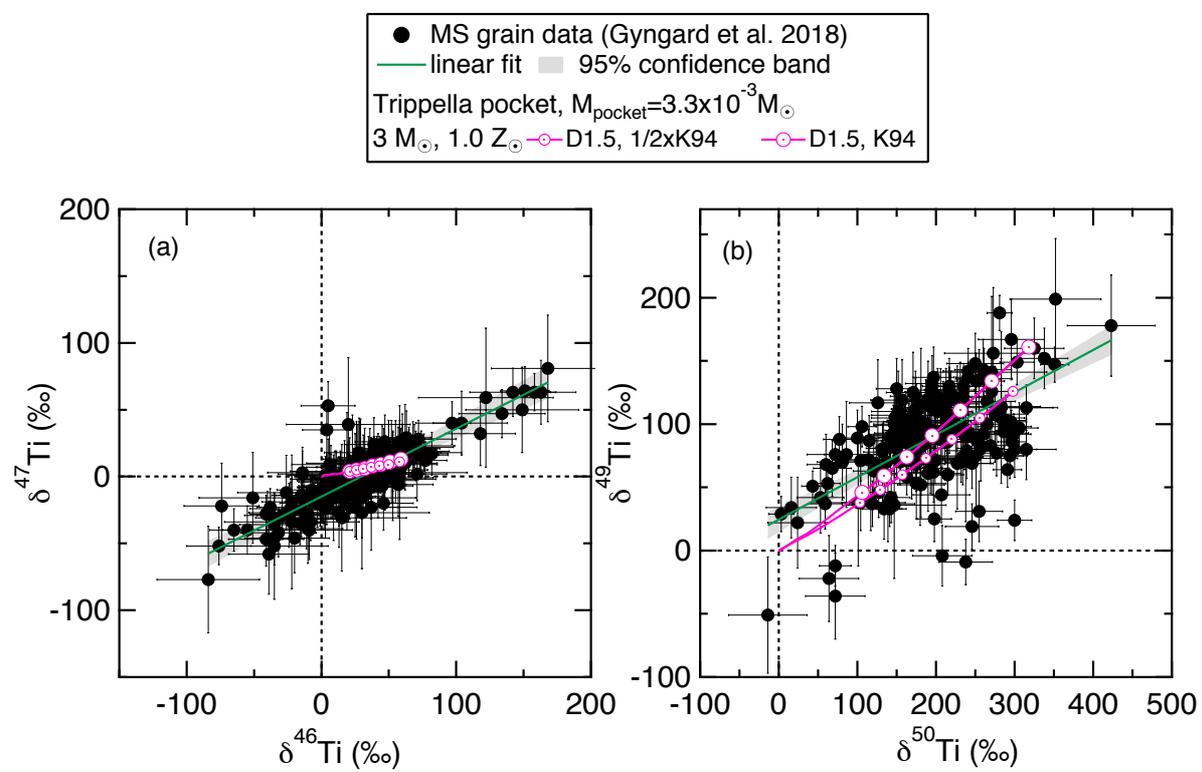